\documentclass{PoS}
\usepackage{amsmath}
\title{EWSB from strongly-coupled dynamics: an EFT approach and implications for $W^+W^-$ production}

\ShortTitle{EWSB from strongly-coupled dynamics: an EFT approach}

\author{\speaker{Oscar Cat\`a}\thanks{This work was performed in the context of the ERC Advanced Grant project 'FLAVOUR' (267104) and was supported in part by the DFG cluster of excellence 'Origin and Structure of the Universe'.}\\
        Ludwig-Maximilians-Universit\"at M\"unchen, Fakult\"at f\"ur Physik,
Arnold Sommerfeld Center for Theoretical Physics, 
D--80333 M\"unchen, Germany
\\
        E-mail: \email{oscar.cata@physik.uni-muenchen.de}}

\abstract{If the dynamics behind EWSB are of strongly-coupled nature, the Standard Model ceases to be renormalizable and should be instead understood as an effective field theory (EFT). Here I will discuss the systematics behind this effective field theory description. My focus will be on deriving a consistent power-counting formula and building the basis of NLO operators. As an application, I will consider $W^+W^-$ production both at linear and hadron colliders. 
}

\FullConference{Xth Quark Confinement and the Hadron Spectrum,\\
		October 8-12, 2012\\
		TUM Campus Garching, Munich, Germany}

\begin{document}
\section{Introduction}
The structure of the Standard Model in the absence of masses is extremely simple. Local symmetry under the gauge group $SU(3)_C\times SU(2)_L\times U(1)_Y$ of the three families of fermions straightforwardly leads to
\begin{align}
{\cal L}_{SM} &=-\frac{1}{2} \langle G_{\mu\nu}G^{\mu\nu}\rangle
-\frac{1}{2}\langle W_{\mu\nu}W^{\mu\nu}\rangle 
-\frac{1}{4} B_{\mu\nu}B^{\mu\nu}+i\sum_j\bar f_j \!\not\!\! Df_j+{\cal{L}}_M
\end{align}
The explicit terms above are completely determined by gauge invariance, which is exact for massless fields. However, fermions and gauge bosons are (with the exception of the photon) massive. Mass terms are generated through the so-called Higgs mechanism, which spontaneously breaks the electroweak gauge invariance. However, it is not clear how the Higgs mechanism is realized in nature. A possibility is Higgs' proposal, namely a linear sigma model with a scalar $SU(2)_L$ doublet satisfying 
\begin{align}
{\cal{L}}_M (\Phi,...)=D_{\mu}\Phi^{\dagger}D^{\mu}\Phi-V(\Phi^{\dagger}\Phi)+{\cal{L}}_{Yukawa}(\Phi),\qquad \Phi=\left(\begin{array}{c}
\phi^+\\ \phi^0
\end{array}\right)
\end{align} 
With the most general renormalizable potential: (i) $\Phi$ can acquire a nontrivial VEV; (ii) the theory is renormalizable; and (iii) as a bonus one gets an accidental global $SU(2)_L\times SU(2)_R$ custodial symmetry. Given the present status of experiments at the LHC~\cite{Aad:2012tfa}, little deviation from this framework seems to be allowed (at least in the gauge boson sector). However, even tiny departures from it would have dramatic effects, {\it{e.g.}} in the unitarization of scattering amplitudes. In order to confirm or disprove the Higgs scenario, it is convenient to adopt a framework where a more flexible implementation of a light scalar (fundamental or not) is possible. This can be achieved if the EWSB is nonlinearly realized. In its minimal version, one assumes the symmetry breaking pattern $SU(2)_L\times SU(2)_R\to SU(2)_V$. The resulting 3 Goldstone modes can be collected in a $SU(2)$ matrix $U$, transforming as $U\rightarrow g_L U g^\dagger_R$, $g_{L,R}\in SU(2)_{L,R}$, whose dynamics is given by the Lagrangian ($L_{\mu}=iUD_{\mu}U^{\dagger}$, $\tau_L=\displaystyle U\frac{\tau_3}{2}U^{\dagger}$):
\begin{equation}\label{lulo}
{\cal L}_M (U,...)= \frac{v^2}{4}\ \langle L_\mu L^\mu\rangle+\beta_1v^2\langle \tau_LL_{\mu}\rangle^2+{\cal{L}}_{Yukawa}(U)
+
\sum_i c_i\frac{v^{6-d_i}}{\Lambda^2}\, {\cal O}_i\nonumber,\qquad U=\exp(i\varphi^a\tau^a/v)
\end{equation}
In this general framework the theory is still renormalizable, but only order by order in the $v/\Lambda$ expansion, which is a consequence of the nondecoupling nature of the new strong sector ($\Lambda\simeq 4\pi v$). Furthermore, custodial symmetry is not built-in, and it is actually broken already at leading order by the second operator above. Due to phenomenological constraints, one typically fine-tunes $\beta_1$ to vanish at tree level. Contributions to $\beta_1$ are then generated by quantum corrections at the one-loop level, which makes $\langle \tau_LL_{\mu}\rangle^2$ a NLO operator. 

Both the linear and nonlinear realization of EWSB implement the Higgs mechanism and thus provide the gauge bosons with masses. The structure of quantum corrections is however different in both scenarios. In order to study their quantum features, one needs a consistent enumeration of operators based on some expansion criteria or power-counting. For the linear case, the power-counting is trivial: operators are simply organized as inverse powers of a cutoff scale. In the nonlinear case, the nondecoupling nature of the interactions makes things a bit more involved and due care has to be exercised. My discussion in this paper will concentrate exclusively on scalar-independent operators in strongly-coupled scenarios, following Ref.~\cite{Buchalla:2012qq}. A light scalar can always be reinstated in the theory by dressing the effective operators with scalar functions and derivatives thereof, {\it{e.g.}},
\begin{eqnarray}
{\cal{O}}_j(U,\psi,W)\to {\cal{O}}_j(U,\psi,W)f_j(\phi),\qquad &
f_j(\phi)=1+a_j^{(1)}\phi+a_j^{(2)}\phi^2+\cdots=\displaystyle\sum_k a_j^{(k)}\phi^k
\end{eqnarray}
This general recipe has been used for instance in~\cite{Contino:2010mh}, though the full systematics of it has not been fully worked out. 

\section{Power-counting and Effective Lagrangian to NLO}
Any EFT requires an organizational principle to classify the operators in terms of the parameter(s) of the series expansion. For strongly-coupled dynamics behind EWSB, the expansion parameter is $v^2/\Lambda^2=1/16\pi^2$. In order to find a consistent power-counting we will only require that the leading-order Lagrangian 
\begin{align}\label{SMLO}
{\cal L}_{SM}={\cal L}_{Kin}+\frac{v^2}{4}\ \langle L_{\mu}L^{\mu}\rangle+{\cal{L}}_{Yukawa}(U)
\end{align}
be homogeneous. Higher order operators will act as counterterms, and accordingly will be loop-generated by the previous Lagrangian. The degree of divergence $D$ of each diagram that one can construct is then given by the master formula~\cite{Buchalla:2012qq}
\begin{equation}\label{master}
D\sim 
\frac{(yv)^{\nu}(gv)^{\xi}}{v^{F_L+F_R-2}} \frac{p^d}{\Lambda^{2L}}
\bigg[\bar\psi_L^{F^1_L} \psi_L^{F^2_L} \bar\psi_R^{F^1_R} \psi_R^{F^2_R}\bigg]\left(\frac{X_{\mu\nu}}{v}\right)^V\left(\frac{\varphi}{v}\right)^B
\end{equation}
where
\begin{equation}\label{powerd}
d\equiv 2L+2-\nu-\frac{F_L+F_R}{2}-V-\xi
\end{equation}
The precise definition of $\nu$ and $\xi$ is given in~\cite{Buchalla:2012qq}. For my purposes here it will suffice to note that $d$ is bounded from above, which makes the power-counting consistent, {\it{i.e.}}, the number of counterterms finite. By repeatedly acting with Eq.~(\ref{master}) on all the independent operators one can construct with the building blocks (gauge bosons, leptons, U field and their derivatives) one concludes~\cite{Buchalla:2012qq} that at NLO there are only 6 classes of operators, to be denoted as $UD^4$, $XUD^2$, $X^2U$, $\psi^2UD$, $\psi^2UD^2$ and $\psi^4U$. Concerning the $\psi^4U$ class, there are (5+11) ${\bar{L}}L{\bar{L}}L$ operators, (7+0) ${\bar{R}}R{\bar{R}}R$, (9+9) ${\bar{L}}L{\bar{R}}R$, (4+8) ${\bar{L}}R{\bar{L}}R$ with global null hypercharge and (0+11) ${\bar{L}}R{\bar{L}}R$ with global hypercharge 1, where the terms in parenthesis count the operators without and with $U$ fields, respectively. The classes $\psi^2UD$ and $\psi^2UD^2$ comprise fermionic single-current operators (vectorial, scalar and tensorial). Their total number is $\{{\cal{O}}_V;{\cal{O}}_S;{\cal{O}}_T\}=\{10;9;6\}$, a sample of which is
\begin{eqnarray}
{\cal O}_{V}^{(1)}=i\bar l\gamma^\mu U P_{22}U^{\dagger} l\ \langle \tau_L L_\mu \rangle ,\quad &
{\cal O}_{V}^{(2)}=i\bar l \gamma^{\mu}UP_{12}U^{\dagger}l\ \langle L_\mu P_{21}\rangle ,\quad &
{\cal O}_{V}^{(3)}=i\bar e\gamma^\mu e \langle \tau_L L_\mu \rangle
\end{eqnarray}
\begin{eqnarray}
{\cal O}_{S}^{(1)}=\bar l UP_{22} \eta\ \langle L_\mu L^\mu \rangle ,\quad & 
{\cal O}_{S}^{(2)}= \bar l UP_{22} \eta\ \langle \tau_L L_\mu \rangle^2 ,\quad &
{\cal O}_{S}^{(3)}= \bar l UP_{12} \eta\ \langle L_\mu P_{21}\rangle \ 
\langle \tau_L L^\mu \rangle
\end{eqnarray}
\begin{eqnarray}
{\cal O}_{T}^{(1)}= \bar l\sigma^{\mu\nu} UP_{12} \eta\ 
\langle L_\mu P_{21}\rangle \langle \tau_L L_\nu \rangle ,\quad &       
{\cal O}_{T}^{(2)}= \bar l\sigma^{\mu\nu} UP_{22} \eta\ 
\langle L_\mu P_{12}\rangle \langle L_\nu P_{21}\rangle          
\end{eqnarray}
where $P_{(11;22)}=\frac{1}{2}(1\pm\tau_3)$ and $P_{(12;21)}=\frac{1}{2}(\tau_1\pm i\tau_2)$.
Finally, the operators without fermions (classes $UD^4$, $XUD^2$ and $X^2U$) are given by~\cite{Longhitano:1980iz}
\begin{eqnarray}\label{pure}
{\cal O}_{D}^{(1)}= \langle L_{\mu}L^{\mu}\rangle \langle L_{\nu}L^{\nu}\rangle,\quad &
{\cal O}_{D}^{(2)}= \langle L_{\mu}L_{\nu} \rangle
  \ \langle L^{\mu}L^{\nu} \rangle,\quad &
{\cal O}_{D}^{(3)}= \langle \tau_L L^{\mu}\rangle^2 \langle \tau_L L^{\nu}\rangle^2  \nonumber\\ 
{\cal O}_{D}^{(4)}= \langle\tau_LL^{\mu}\rangle \langle \tau_LL_{\mu}\rangle \langle L_{\nu}L^{\nu}\rangle,\quad &
{\cal O}_{D}^{(5)}= \langle\tau_LL^{\mu}\rangle \langle \tau_LL_{\nu}\rangle \langle L_{\mu}L^{\nu}\rangle
\end{eqnarray}
and~\cite{Appelquist:1993ka}
\begin{align}\label{naive1}
{\cal{O}}_{XU}^{(1)}&=g^{\prime}gB_{\mu\nu}\langle W^{\mu\nu}\tau_L\rangle\qquad &&{\cal{O}}_{XU}^{(4)}=g^{\prime}g\epsilon_{\mu\nu\lambda\rho}\langle \tau_L W^{\mu\nu}\rangle 
B^{\lambda\rho}\nonumber\\
{\cal{O}}_{XU}^{(2)}&=g^2 \langle W^{\mu\nu}\tau_L\rangle^2 \qquad&&{\cal{O}}_{XU}^{(5)}=g^2\epsilon_{\mu\nu\lambda\rho}\langle\tau_LW^{\mu\nu}\rangle
\langle\tau_LW^{\lambda\rho}\rangle\nonumber\\
{\cal{O}}_{XU}^{(3)}&=g\epsilon^{\mu\nu\lambda\rho}\langle W_{\mu\nu}L_{\lambda}\rangle\langle\tau_L L_{\rho}\rangle\qquad &&{\cal{O}}_{XU}^{(6)}=g\langle W_{\mu\nu}L^{\mu}\rangle\langle \tau_LL^{\nu}\rangle
\nonumber\\
{\cal{O}}_{XU}^{(7)}&=ig^{\prime}B_{\mu\nu}\langle\tau_L[L^{\mu},L^{\nu}]
\rangle\qquad &&{\cal{O}}_{XU}^{(10)}=ig^{\prime}\epsilon_{\mu\nu\lambda\rho}B^{\mu\nu}\langle\tau_L[L^{\lambda},L^{\rho}]
\rangle\nonumber\\
{\cal{O}}_{XU}^{(8)}&=ig\langle W_{\mu\nu}[L^{\mu},L^{\nu}]
\rangle\qquad &&{\cal{O}}_{XU}^{(11)}=ig\epsilon_{\mu\nu\lambda\rho}\langle W^{\mu\nu}[L^{\lambda},L^{\rho}]
\rangle\nonumber\\
{\cal{O}}_{XU}^{(9)}&=ig\langle W_{\mu\nu}\tau_L\rangle \langle \tau_L[L^{\mu},L^{\nu}]
\rangle\qquad &&{\cal{O}}_{XU}^{(12)}=ig\epsilon_{\mu\nu\lambda\rho}\langle W^{\mu\nu}\tau_L\rangle \langle \tau_L[L^{\lambda},L^{\rho}]
\rangle
\end{align}
Form a phenomenological viewpoint, the operators in Eq.~(\ref{pure}) correspond to anomalous quartic gauge couplings. In the unitary gauge they take the form
\begin{equation}
{\cal O}_{D}\sim\bigg\{Z_{\mu}Z^{\mu}Z_{\nu}Z^{\nu};~W_{\mu}^+W^{+\mu}W_{\nu}^-W^{-\nu};~W_{\mu}^+W^{-\mu}W_{\nu}^+W^{-\nu};~ Z_{\mu}Z^{\mu}W_{\nu}^+W^{-\nu};~Z^{\mu}Z^{\nu}W_{\nu}^+W_{\mu}^-\bigg\}
\end{equation}
which indeed exhausts all the possible quartic contractions of gauge bosons. Eq.~(\ref{naive1}) instead collects the CP-even (left column) and CP-odd (right column) operators responsible for oblique and triple gauge corrections. As a matter of fact, only half the operators in Eq.~(\ref{naive1}) are independent. By using the equations of motion for the gauge fields 
\begin{eqnarray}
\partial^{\mu}B_{\mu\nu}=g^{\prime}\left[Y_j{\bar{f}}_j\gamma_{\nu}f_j+\frac{v^2}{2}\langle \tau_LL_{\nu}\rangle\right];\qquad &
D^{\mu}W_{\mu\nu}^a=\displaystyle\frac{g}{2}\left[{\bar{f}}_{jL}\gamma_{\nu}\tau^a f_{jL}-\frac{v^2}{2}\langle \tau^aL_{\nu}\rangle\right]
\end{eqnarray}
and the identities
\begin{eqnarray}
D_{\mu}\tau_L=-i[\tau_L,L_{\mu}];\qquad &
D_{[\mu}L_{\nu]}=gW_{\mu\nu}-g^{\prime}B_{\mu\nu}\tau_L+i[L_{\mu},L_{\nu}]
\end{eqnarray}
one can show that
\begin{eqnarray}\label{relations}
&{\cal{O}}_{XU}^{(7)}=f_7({\cal{O}}_{XU}^{(1)},{\cal{O}}_{V}^{(j)},{\cal{O}}_{\beta});\,\quad  {\cal{O}}_{XU}^{(8)}=f_8({\cal{O}}_{XU}^{(1)},{\cal{O}}_{V}^{(j)});\,\quad 
{\cal{O}}_{XU}^{(9)}=f_9({\cal{O}}_{XU}^{(2)},{\cal{O}}_{V}^{(j)},{\cal{O}}_{\beta})\nonumber\\
&{\cal{O}}_{XU}^{(10)}=-{\cal{O}}_{XU}^{(4)};\,\qquad\qquad\qquad {\cal{O}}_{XU}^{(11)}=-{\cal{O}}_{XU}^{(4)};\qquad\qquad\quad
{\cal{O}}_{XU}^{(12)}=-\frac{1}{2}{\cal{O}}_{XU}^{(5)}
\end{eqnarray}
These relations were noticed before~\cite{De Rujula:1991se} but their role in phenomenology was never exploited. Yet they are of importance, as I will show below for $W^+W^-$ production.
 
The 6 classes of operators outlined above constitute the most general description of leading new physics effects at low energies. Bits of it were worked out for the last 30 years~\cite{Appelquist:1984rr}. However, a full systematic treatment, {\it{i.e.}}, providing (i) a well-defined power-counting; (ii) a complete basis of operators; and (iii) free from redundancies, was absent in the literature. These ingredients are essential to perform consistent analyses of electroweak data. 

\section{$W^+W^-$ production at linear and hadron colliders}
As an illustrative example of the potential applications of the EFT developed in the previous Section I will consider $W^+W^-$ production, which has been one of the benchmark processes in the study of anomalous triple gauge vertices (TGVs). For simplicity I will discuss $W^+W^-$ production at linear colliders, which already captures the main qualitative features I want to illustrate. In what follows I will stick rather closely to the analysis of Ref.~\cite{Buchalla:2013wpa}. Comments on $W^+W^-$ at hadron colliders will be given at the end of the Section. For a discussion of $ZZ$ and $\gamma Z$ production, the reader is referred to Ref.~\cite{Cata:2013sva}. 

$e^+e^-\to W^+W^-$ in the Standard Model can proceed through $e^+e^-$ annihilation or $e^+e^-$ exchange, whose contributions can be extracted from Eq.~(\ref{SMLO}). New physics corrections to these results are parametrized in full generality by the following subset of NLO operators:{\footnote{${\cal{O}}_W=g^3\varepsilon_{abc}W_{\mu}^{a\nu}W_{\nu}^{b\rho}W_{\rho}^{c\mu}$ and ${\tilde{{\cal{O}}}}_W=g^3\varepsilon_{abc}{\tilde{W}}_{\mu}^{a\nu}W_{\nu}^{b\rho}W_{\rho}^{c\mu}$ can be actually shown to be NNLO in both the linear and nonlinear realization of EWSB. However, it will prove instructive to keep them all through our analysis.}
\begin{align}
{\cal{L}}_{NLO}&=\sum_{j=1}^{6}\lambda_j{\cal{O}}_{XUj}+\sum_{j=1}^3 \eta_j{\cal{O}}_{V}^{(j)}+\beta_1{\cal{O}}_{\beta}+a_W{\cal{O}}_W+{\tilde{a}}_W{\tilde{{\cal{O}}}}_W
\end{align}
which correct the SM gauge-fermion vertices ($e^+e^-Z$ and $\nu_ee^{\pm}W^{\mp}$) and the triple gauge vertices ($W^+W^-Z$ and $W^+W^-\gamma$), but also shift the photon and $Z$ propagators (through the oblique ${\cal{O}}_{XU1,2}$) and the electroweak parameter triad $(\alpha,m_Z,G_F)$. It is convenient to reabsorb the shifts in propagators and EW parameters by the 2-step procedure described in~\cite{Holdom:1990xq}:
\begin{itemize}
\item[(1)] Canonical normalization of the kinetic terms through the following field redefinitions:
\begin{eqnarray}
Z_{\mu}\to(1+e^2\Delta_Z)~Z_{\mu}^{\prime},\qquad &A_{\mu}\to(1+e^2\Delta_A)~A_{\mu}^{\prime}+e^2\Delta_{AZ}~Z_{\mu}^{\prime}
\end{eqnarray}
where
\begin{eqnarray}
\Delta_Z=-\lambda_1+\frac{\lambda_2}{2 t_W^2};\,\qquad &
\Delta_A=\displaystyle\lambda_1+\frac{\lambda_2}{2};\,\qquad &
\Delta_{AZ}=\frac{2\lambda_1}{t_{2W}}+\frac{\lambda_2}{t_W}
\end{eqnarray}
\item[(2)] Renormalization of the Standard Model parameters ($\alpha$, $m_Z$, $G_F$) through
\begin{equation}
e\to(1-e^2\Delta_A)~e^{\prime};\quad m_Z\to(1-e^2\Delta_Z+\beta)~m_Z^{\prime};\quad
s_{W}\to(1-\xi)~s_W^{\prime};\quad c_{W}\to(1+t_W^2\xi)~c_W^{\prime}
\end{equation}
where 
\begin{align}
\xi&=\frac{c_W^2}{c_W^2-s_W^2}(e^2\Delta_A-e^2\Delta_Z+\beta_1-2\eta_{2})
\end{align} 
\end{itemize}
Once this is done, the new physics corrections affect only the gauge-fermion and triple gauge vertices, which can be parametrized in full generality by
\begin{align}
{\cal{L}}_f&=e{\bar{f}}\gamma^{\mu}A_{\mu}f+e\sum_{j=L,R}{\zeta}_j{\bar{f}}_j\gamma^{\mu}Z_{\mu}f_j-\frac{g}{\sqrt{2}}\phi_L{\bar{\nu}}_L\gamma^{\mu}W_{\mu}^+f_L+{\mathrm{h.c.}}\nonumber\\
\frac{1}{e}{\cal{L}}_{TGV}&=i\kappa_V W^+_{\mu}W^-_{\nu}V^{\mu\nu}+ig_{1V}(W_{\mu\nu}^+W^{\mu-}-W_{\mu\nu}^-W^{\mu+})V^{\nu}+i\frac{\lambda_V}{\Lambda^2}W^{+\nu}_{\mu}W^{-}_{\nu\lambda}V^{\lambda\mu}\nonumber\\
&+g_{4V}(W_{\mu\nu}^+W^{\mu-}+W_{\mu\nu}^-W^{\mu+})V^{\nu}-g_{5V}({\tilde{W}}_{\mu\nu}^+W^{\mu-}+{\tilde{W}}_{\mu\nu}^-W^{\mu+})V^{\nu}\nonumber\\
&+i{\tilde{\kappa}}_VW^+_{\mu}W^-_{\nu}{\tilde{V}}^{\mu\nu}+i\frac{{\tilde{\lambda}}_V}{\Lambda^2}W^{+\nu}_{\mu}W^{-}_{\nu\lambda}{\tilde{V}}^{\lambda\mu}
\end{align}
The triple-gauge and gauge-fermion coefficients above can be generically expressed as $\rho_{iV}=\rho_{iV}^{(0)}+e^2\delta\rho_{iV}$, where the first piece collects the SM contribution, which is nonvanishing for
\begin{eqnarray}
\zeta_L^{(0)}&=t_{2W}^{-1};\qquad  \kappa_Z^{(0)}&=g_{1Z}^{(0)}=-t_W^{-1};\qquad \phi_L=1;\nonumber\\
\zeta_R^{(0)}&=-t_{W};\qquad  \kappa_A^{(0)}&=g_{1A}^{(0)}=-1
\end{eqnarray}
while $\delta\rho_{iV}$ contains the new physics corrections. In the EFT language we want to adopt here, $\delta\rho_{iV}=f_{iV}(\lambda_j,\eta_j,\beta_1,a_W,{\tilde{a}}_W)$. For the time being, however, we will keep their dependence implicit.

The Feynman rule for the gauge-fermion vertex is trivial, while for the triple-gauge vertex one finds~\cite{Hagiwara:1986vm}:
\begin{align}\label{feyn}
\frac{1}{e}\Gamma^{VW^+W^-}_{\mu\lambda\nu}(q,p^+,&p^-)=-i(\kappa_V+\lambda_V+g_1^V)[q_{\lambda}g_{\mu\nu}-q_{\nu}g_{\mu\lambda}]-i\left(g_1^V+\frac{\lambda_V}{2}\frac{s}{\Lambda^2}\right)[(p^+-p^-)_{\mu}g_{\nu\lambda}]\nonumber\\
&+i\frac{\lambda_V}{\Lambda^2}\left[(p^+-p^-)_{\mu}q_{\nu}q_{\lambda}\right]-g_4^V[q_{\lambda}g_{\mu\nu}+q_{\nu}g_{\mu\lambda}]-g_5^V(p^+-p^-)^{\rho}\epsilon_{\mu\nu\rho\lambda}\nonumber\\
&+i({\tilde{\kappa}}_V+{\tilde{\lambda}}_V)\epsilon_{\mu\nu\rho\lambda}q^{\rho}+i\frac{{\tilde{\lambda}}_V}{\Lambda^2}\left[\frac{1}{2}(p^+-p^-)_{\mu}\epsilon_{\nu\lambda\rho\sigma}q^{\rho}(p^+-p^-)^{\sigma}\right]
\end{align}
In previous analyses of $W^+W^-$ production it has been common to neglect the gauge-fermion vertex corrections and work with the triple vertex corrections alone, assuming that they satisfy a dipole structure. Such a strategy has some fundamental deficiencies. First, since gauge-fermion and triple-gauge operators are related by the equations of motion, neglecting gauge-fermion operators altogether violates fundamental field theoretical relations. Second, the different triple-gauge coefficients are not independent but correlated by the underlying $SU(2)_L\times U(1)_Y$ symmetry, to which the dipole parametrization is blind. Since the dipole approximation does not respect gauge symmetry it can generate fake violations of unitarity that have nothing to do with new physics. In order to illustrate these drawbacks, let us consider the leading effects in the cross sections for unpolarized $WW$ pairs, {\it{i.e.}}, linear corrections in the new physics parameters in the large-$s$ limit:     
\begin{align}\label{result}
\frac{d\sigma(e^-_R e^+_L\to W^-W^+)}{d\cos\theta}&=\frac{s_{\theta}^2e^4}{64\pi m_Z^2s_Wc_W^5}\bigg[-c_W^2\delta \zeta_R+e^2s_W(c_W\delta \kappa_A-s_W\delta \kappa_Z)\bigg]\nonumber\\
\frac{d\sigma(e^-_L e^+_R\to W^-W^+)}{d\cos\theta}&=\frac{s_{\theta}^2e^4}{256\pi m_Z^2s_W^4c_W^5}\bigg[2c_W(\delta\phi_L-s_Wc_W\delta\zeta_L)+e^2s_W(s_{2W}\delta\kappa_A-c_{2W}\delta\kappa_Z)\bigg]
\end{align}
First of all, notice that the $\lambda_{V}$ coefficients are absent, even though they seem to appear $s$-enhanced in Eq.~(\ref{feyn}). This is precisely because of $SU(2)\times U(1)$-induced cancellations, which are completely obliterated by a naive dipole ansatz. Second, the presence of gauge-fermion operators is fundamental. Actually, without them the expressions above would vanish. This can be explicitly checked by substituting $\delta \kappa_{A}, \delta \kappa_{Z}, \delta\zeta_L, \delta\zeta_R, \delta\phi_L$ in terms of the EFT coefficients. However, it is more enlightening to rederive the results in the Landau gauge with the help of the equivalence theorem. This states that the most divergent contributions to $WW$ production should come from longitudinally-polarized $W$'s, {\it{i.e.}}, from $e^+e^-\to \varphi^+\varphi^-$.

The calculation in that case turns out to be very simple~\cite{Buchalla:2013wpa}. The SM only contributes to the $s$-channel, with the $(\gamma,Z)\varphi^+\varphi^-$ vertices coming from the Goldstone kinetic term. New physics contributions can instead be shown to be purely local, coming entirely from the gauge-fermion operators. 
The interference between the Standard Model and the new physics contribution can be easily computed and results in
\begin{align}\label{res1}
\frac{d\sigma(e^-_R e^+_L\to W^-W^+)}{d\cos\theta}&=\frac{\pi \alpha^2 \sin^2\theta}{8 s_W^2c_W^2}\frac{1}{m_W^2}\eta_3\nonumber\\
\frac{d\sigma(e^-_L e^+_R\to W^-W^+)}{d\cos\theta}&=\frac{\pi \alpha^2 \sin^2\theta}{16 c_W^2s_W^4}\frac{1}{m_W^2}\left(\eta_{1}+2\eta_{2}\right)
\end{align}
Direct substitution in Eqs.~(\ref{result}) would have delivered the same result, but through intricate cancellations that would have obscured the physics. Gauge-fermion operators are the leading contribution because they are the only NLO operators that contribute to $e^+e^-\to \varphi^+\varphi^-$.   

It is instructive at this point to unfold the relations between gauge-fermion, oblique and triple-gauge operators of Eqs.~(\ref{relations}) and express the previous results in terms of triple-gauge operators. The results then take the form
\begin{align}\label{res2}
\frac{d\sigma(e^-_R e^+_L\to W^-W^+)}{d\cos\theta}&=-\frac{\pi^2 \alpha^3 \sin^2\theta}{ s_W^2c_W^4}\frac{1}{m_W^2}\lambda_{7}\nonumber\\
\frac{d\sigma(e^-_L e^+_R\to W^-W^+)}{d\cos\theta}&=-\frac{\pi^2 \alpha^3 \sin^2\theta}{4 c_W^4s_W^6}\frac{1}{m_W^2}\left(s_W^2\lambda_{7}+c_W^2\left(\lambda_{8}+\frac{1}{2}\lambda_{9}\right)\right)
\end{align}
Comparing Eqs.~(\ref{res1}) and (\ref{res2}) above, the change of basis is effected by
\begin{eqnarray}
\lambda_7=-\frac{c_W^2}{8\pi\alpha}\eta_{3};~&
\lambda_8=\displaystyle\frac{s_W^2}{4\pi\alpha}\left(\eta_{1}-\frac{1}{2}\eta_{3}\right);~&
\lambda_9=-\frac{s_W^2}{\pi\alpha}\left(\eta_{1}+\eta_{2}-\frac{1}{2}\eta_{3}\right)
\end{eqnarray}
At first sight, it might seem that these relations are at odds with Eqs.~(\ref{relations}). Note however that Eqs.~(\ref{relations}) hold for any value of the energy. What we have found above instead is their large-$s$ limit, which simplifies them notably: in the high-energy limit Eqs.~(\ref{relations}) 'project out' to  
\begin{eqnarray}
{\cal{O}}_{XU}^{(7)}\stackrel{(s\to \infty)}{=}f_7({\cal{O}}_{V}^{(1)},{\cal{O}}_{V}^{(3)});\,\quad &{\cal{O}}_{XU}^{(8)}\stackrel{(s\to \infty)}{=}f_8({\cal{O}}_{V}^{(1)},{\cal{O}}_{V}^{(2)});\,\quad &
{\cal{O}}_{XU}^{(9)}\stackrel{(s\to \infty)}{=}f_9({\cal{O}}_{V}^{(2)})
\end{eqnarray}
\begin{figure}[t]
\begin{center}
\includegraphics[width=4.7cm]{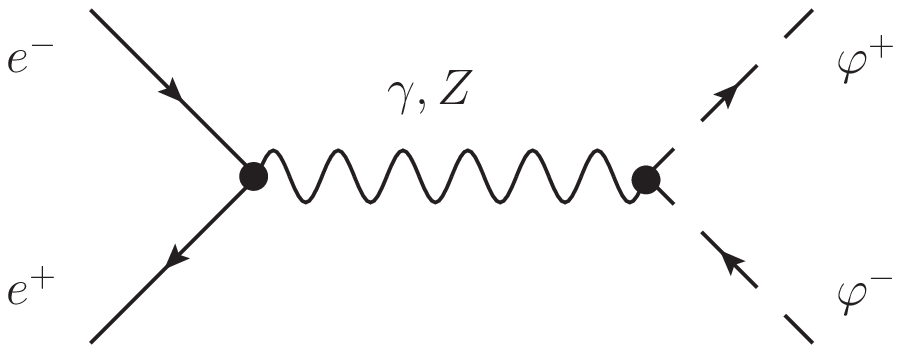}\hskip 0.5cm
\includegraphics[width=3.0cm]{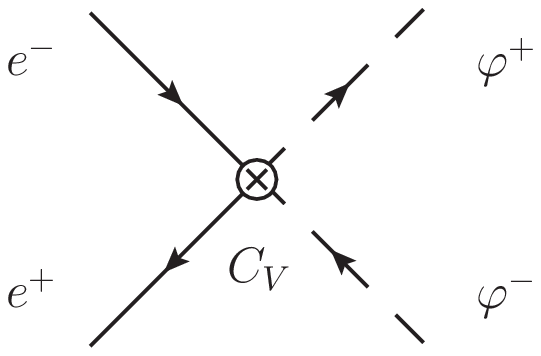}\hskip 0.5cm
\includegraphics[width=4.7cm]{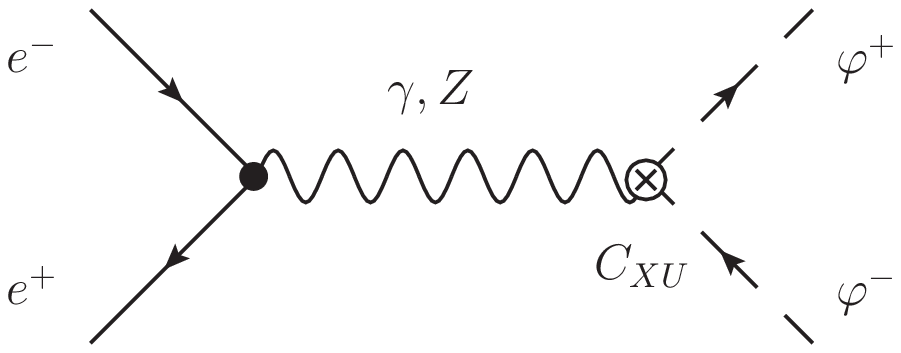}
\end{center}
\caption{\small{\it{Different contributions to $e^+e^-\to \varphi^+\varphi^-$. From left to right: (i) Standard Model piece; new physics contribution in terms of (ii) gauge-fermion operators and (iii) triple-gauge operators.}}}\label{fig:4}
\end{figure} 
So far I have been discussing $W^+W^-$ production at linear colliders. At hadron colliders the calculations are more involved due to hadronization, but the qualitative picture remains. At the partonic level, the number of gauge-fermion operators gets doubled and, following the arguments above, one can conclude that 5 of them will provide the leading new physics effects in $pp\to W^+W^-$. In order to be quantitative, their coefficients would have to be weighted by PDF's. Work in this direction is currently underway and should provide a consistent framework for new physics searches in $W^+W^-$ production at the LHC.
 
\section{Conclusions}

The main conclusions one can extract from our analysis of $W$ pair production can be summarized in the following points:
\begin{itemize}
\item A form factor analysis with a dipole ansatz for the triple gauge vertices (TGVs) is in general inconsistent with gauge symmetry and can thus fake violations of unitarity. The only way to guarantee field-theoretical consistency is to work with a full-fledged EFT, which is the most general field theory at a given scale. In particular, an EFT analysis shows that the TGV parameters $\lambda_{Z,\gamma}$, which naively would be $s$-enhanced, are actually strongly suppressed due to $SU(2)_L\times U(1)_Y$-induced cancellations.  
\item $W^+W^-$ production is, strictly speaking, not a probe of anomalous TGVs, as commonly stated. Gauge-fermion vertices are equally important and cannot be neglected. Actually, for $e^+e^-\to W^+W^-$ one can describe the leading new physics effects entirely in terms of gauge-vertex operators or gauge-fermion ones. Both descriptions happen to be dual. Therefore, in a phenomenological fit one does not need to neglect gauge-fermion operators: they can be eliminated from the picture altogether.   
\item $e^+e^-\to W^+W^-$ has the peculiarity that one can trade the 3 gauge-fermion operators for triple-gauge operators and vice versa but in $pp\to W^+W^-$, for instance, this is no longer the case. Therefore, given that the number of gauge-fermion operators at NLO is much bigger than that of triple-gauge operators, it seems more natural to eliminate the latter, especially in view of fits involving multiple processes. 
\end{itemize}


\begin{thebibliography}{99}

\bibitem{Aad:2012tfa} 
  G.~Aad {\it et al.}  [ATLAS Collaboration],
  Phys.\ Lett.\ B {\bf 716}, 1 (2012)
  [arXiv:1207.7214 [hep-ex]];
  S.~Chatrchyan {\it et al.}  [CMS Collaboration],
  Phys.\ Lett.\ B {\bf 716}, 30 (2012)
  [arXiv:1207.7235 [hep-ex]].

\bibitem{Buchalla:2012qq} 
  G.~Buchalla and O.~Cata,
  JHEP {\bf 1207}, 101 (2012)
  [arXiv:1203.6510 [hep-ph]].
 
\bibitem{Contino:2010mh} 
  R.~Contino, C.~Grojean, M.~Moretti, F.~Piccinini and R.~Rattazzi,
  JHEP {\bf 1005}, 089 (2010)
  [arXiv:1002.1011 [hep-ph]].

\bibitem{Longhitano:1980iz}
  A.~C.~Longhitano,
  Phys.\ Rev.\  D {\bf 22}, 1166 (1980).

\bibitem{Appelquist:1993ka}
  T.~Appelquist and G.~H.~Wu,
  Phys.\ Rev.\  D {\bf 48}, 3235 (1993)
  [arXiv:hep-ph/9304240].
      
\bibitem{De Rujula:1991se} 
  A.~De Rujula, M.~B.~Gavela, P.~Hernandez and E.~Masso,
  Nucl.\ Phys.\ B {\bf 384}, 3 (1992).
  A.~Nyffeler and A.~Schenk,
  Phys.\ Rev.\ D {\bf 62}, 113006 (2000)
  [hep-ph/9907294].
  C.~Grojean, W.~Skiba and J.~Terning,
  Phys.\ Rev.\ D {\bf 73}, 075008 (2006)
  [hep-ph/0602154].

\bibitem{Appelquist:1984rr}
  T.~Appelquist, M.~J.~Bowick, E.~Cohler and A.~I.~Hauser,
  Phys.\ Rev.\  D {\bf 31}, 1676 (1985).
  R.~D.~Peccei and X.~Zhang,
  Nucl.\ Phys.\  B {\bf 337}, 269 (1990).
  E.~Bagan, D.~Espriu and J.~Manzano,
  Phys.\ Rev.\  D {\bf 60}, 114035 (1999)
  [arXiv:hep-ph/9809237].

\bibitem{Buchalla:2013wpa} 
  G.~Buchalla, O.~Cata, R.~Rahn and M.~Schlaffer,
  arXiv:1302.6481 [hep-ph].
  
\bibitem{Cata:2013sva} 
  O.~Cata,
  arXiv:1304.1008 [hep-ph].
  
\bibitem{Holdom:1990xq}
  B.~Holdom,
  Phys.\ Lett.\  B {\bf 258}, 156 (1991).

\bibitem{Hagiwara:1986vm}
  K.~Hagiwara, R.~D.~Peccei, D.~Zeppenfeld and K.~Hikasa,
  Nucl.\ Phys.\  B {\bf 282}, 253 (1987).

\end{thebibliography}
\end{document}